\documentclass[twocolumn,showpacs,aps,prl]{revtex4}
\usepackage{graphicx}
\usepackage{dcolumn}
\usepackage{amsmath}

\newcommand{ \rts }{ \sqrt{s_{\rm NN}} }
\newcommand{ \pb  }{ \overline{p} }

\newcommand{ \pbp }{ \overline{p}/p }

\begin{document}

\title{ Mid-rapidity anti-proton to proton ratio from Au+Au collisions
  at $ \sqrt{s_{\rm NN}} = 130$ GeV }

\author{
C.~Adler$^{11}$, Z.~Ahammed$^{25}$, C.~Allgower$^{12}$, M.~Anderson$^5$, 
G.S.~Averichev$^{9}$, J.~Balewski$^{12}$, O.~Barannikova$^{9,25}$, 
L.S.~Barnby$^{15}$, J.~Baudot$^{13}$, S.~Bekele$^{22}$
% ...
}
\address{
% ...
$^5$University of California, Davis, California 95616 \\
% ...
$^{9}$Laboratory for High Energy (JINR), Dubna, Russia \\
% ...
$^{11}$University of Frankfurt, Frankfurt, Germany \\
$^{12}$Indiana University, Bloomington, Indiana 47408 \\
$^{13}$Institut de Recherches Subatomiques, Strasbourg, France \\
% ...
$^{15}$Kent State University, Kent, Ohio 44242 \\
% ...
$^{22}$Ohio State University, Columbus, Ohio 43210 \\
% ...
$^{25}$Purdue University, West Lafayette, Indiana 47907 \\
% ...
}

\date{\today}

\begin{abstract}
  We report results on the ratio of mid-rapidity anti-proton to proton
  yields in Au+Au collisions at $\rts = 130$ GeV per nucleon pair as
  measured by the STAR experiment at RHIC. Within the rapidity and
  transverse momentum range of $|y|<0.5$ and 0.4 $<p_t<$ 1.0 GeV/$c$,
  the ratio is essentially independent of either transverse momentum
  or rapidity, with an average of $0.65\pm 0.01_{\rm (stat.)} \pm
  0.07_{\rm (syst.)}$ for minimum bias collisions.  Within errors,
  no strong centrality dependence is observed.  The results indicate
  that at this RHIC energy, although the $p$-$\pb$ pair production
  becomes important at mid-rapidity, a significant excess of baryons
  over anti-baryons is still present.
\end{abstract}

\pacs{25.75.Ld}

\maketitle

Lattice Quantum Chromodynamics calculations predict that at
sufficiently high energy density a phase transition from hadronic
matter to a state of deconfined quarks and
gluons~\cite{qm99,karsch0200} will occur.  To create and study this
deconfined state is the primary goal of the heavy-ion collision
program at the Relativistic Heavy Ion Collider (RHIC)~\cite{qm99}.
However, the formation of such a deconfined state depends on the
initial conditions of the matter created at the early stage of
heavy-ion collisions.  Baryon number transport (or stopping), achieved
mostly at the early stage, is one of the important observables
\cite{busza,vance99} for high-energy collisions, since the degree of
baryon stopping affects the overall dynamical evolution of these
collisions. It affects the processes of initial parton equilibration
\cite{wang92,gyulassy94,baier95,wang98}, particle production
\cite{qm99}, thermal and/or chemical equilibration~\cite{pbm99}, and
the development of collective expansion \cite{na44slope,starflow1}.
Information on baryon transport may be experimentally accessed by the
measurement of the ratio of the anti-proton to proton yields ($\pbp$).

In this letter, we report results on the inclusive $\pbp$ ratio in Au+Au
collisions at the center of mass energy $\rts=130$ GeV per nucleon
pair measured by the Solenoidal Tracker at RHIC (STAR).

The STAR detector consists of several detector sub-systems in a large
solenoidal magnet.  The STAR Time Projection Chamber (TPC)~\cite{TPC}
is 4.2 meters long, with a 50 cm inner radius and a 2 meter outer
radius.  For minimum ionizing particles in P-10 (90\% argon, 10\%
methane) gas, approximately 45 primary electrons are produced per
centimeter of track.  This ionization was measured on 45 pad rows. The
system could accommodate ionization from 200 MeV/c protons without
saturation and its noise level was about 5\% of minimum ionizing. More
details of the detector can be found elsewhere~\cite{fee,star}.  For
the data taken in the year 2000 run and presented here, the main setup
consists of the TPC, a scintillator trigger barrel (CTB) surrounding
the TPC, and two Zero Degree Calorimeters \cite{zdc} located up- and
down-stream along the axis of the TPC and beams. The TPC was operated
in a 0.25 Tesla magnetic field. The CTB measures the total energy
deposition in the scintillator from charged particles, and the ZDCs
measure beam-like neutrons from fragmentation and/or evaporation of
the colliding nuclei.  The coincidence of the ZDCs and RHIC beam
crossing was the experimental trigger for the minimum bias events used
in this analysis.

The collision centrality was determined off-line from the measured
total charged particle multiplicity in the pseudorapidity range of
$|\eta|<0.75$.  As was done in \cite{starflow1}, the
total charged multiplicity was scaled by the maximum value measured,
and the distribution of the scaled multiplicity was subdivided into
eight centrality bins.

\begin{figure}[h]
\centering\mbox{
\includegraphics[width=.48\textwidth]{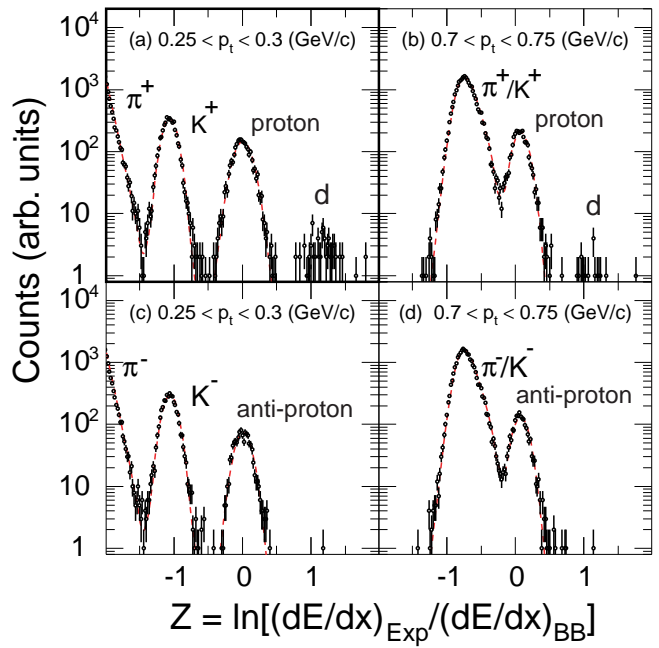}}
\caption{Mid-rapidity particle yields as a function of 
  $Z = log[(dE/dx)_{Exp}/(dE/dx)_{BB}]$ from minimum bias Au+Au
  events.  Top two plots (a)-(b) and bottom plots (c)-(d) are for
  positive particles and negative particles, respectively.  Proton
  mass was used in the Bethe-Bloch $(dE/dx)_{BB}$ calculations so the
  Z-distributions of proton are centered about zero. Dashed-lines are
  the multi-Gaussian fits to the measured distributions.}
\label{dedx}
\end{figure}

For this analysis, 68k minimum bias events were used with an event
vertex $|z| \le$ 50 cm. Protons and anti-protons were selected
according to specific energy loss ($dE/dx$) in the TPC up to a
momentum of 1 GeV/$c$. A mean $dE/dx$ for each track was obtained by
averaging the lower 70\% of the measured $dE/dx$ values.  This
selection reduces our sensitivity to large fluctuations in the $dE/dx$
measurements.  At a momentum of 0.5 GeV/$c$, the $dE/dx$ resolution
was found to be 8\% for a typical long track in the TPC. Figure
~\ref{dedx} shows the mid-rapidity ($|y| < 0.1$) particle raw yields
as a function of $Z = log[(dE/dx)_{Exp}/(dE/dx)_{BB}]$, where
$(dE/dx)_{BB}$ is the expected $dE/dx$ value. The proton mass was used
in the Bethe-Bloch $(dE/dx)_{BB}$ calculations so the Z-distributions
of proton (anti-proton) are centered about zero in those plots.  The
dashed-lines are the multi-Gaussian fits to the Z-distributions
(including pions, kaons, and protons).  Protons and anti-protons with
$p_t>175$~MeV/$c$ were reconstructed.  Within the acceptance the
typical momentum resolution was 2\% and the systematic uncertainty in
the proton (anti-proton) yields, extracted from the $dE/dx$ fitting,
was less than 2\%. Tracks were selected if their distance of closest
approach to the primary vertex were less than 3 cm, and if they had at
least 15 out of 45 possible TPC space-points.  Several tests showed
that the $\pbp$ results were not sensitive to variations of these cuts
within reasonable ranges.  The raw yields of protons and anti-protons
were obtained by fitting the $Z$ distributions (see Figure 1) for each
given $(y, p_t)$ bin, and were used to obtain the $\pbp$ ratio.

The STAR TPC is symmetric about mid-rapidity and has full azimuthal
coverage. Due to this symmetry, most of the detector effects are the
same for protons and anti-protons and cancel in the $\pbp$ ratio.
However, there are two important differences which are mostly dominant
at low momentum: (i) background protons are produced in the detector
materials via hadronic interactions~\cite{Schiffer}, and (ii) some of
the anti-protons are absorbed in the detector materials.

Background protons are evident from the distribution of the distance
of closest approach ($dca$).  The distribution is peaked at small $dca$
and has a flat tail from secondary production which extends in to the
peak region where primary tracks are centered.  Since the background
shape of the $dca$ distribution at small $dca$ ($\le 3$ cm) is not
accessible from the data, a Monte Carlo (MC) simulation (GEANT
\cite{geant}) was employed using as input particles generated by both
HIJING \cite{hijingpr} and RQMD \cite{sorge95} models.  An
empirical form was found for the background:

\begin{equation}
{\rm background}(dca) \propto 
\left[1+\exp\left(\frac{dca-dca_0}{a}\right)\right]^{-1}
\label{pback}
\end{equation}
where the parameters $dca_0$ and $a$ were obtained by fitting to the
MC results. In general these parameters are momentum dependent and the
normalization factor was obtained from the data.  In the momentum
region around 0.2 GeV/$c$, the number of background protons was found
to be twice the number of primary protons.  This leads to a rather
large systematic uncertainty in the raw proton yield.  Therefore, in
this analysis we limited ourselves to the region where systematic
errors are below 10\% leading to a lower $p_t$ cut-off at 0.4 GeV/$c$.
At the high momentum end, $p \approx$ 1 GeV/$c$, the $dE/dx$ method
becomes insufficient for particle identification.

The anti-proton absorption loss was estimated via:
\begin{equation}
{\rm absorption} = 1-\exp(-\sigma_{\rm anni} \rho_t p/p_t),
\label{anni}
\end{equation}
where $\rho_t$ is the transverse area density of nucleons in the
materials. $p$ and $p_t$ are the anti-proton total momentum and
transverse momentum, respectively. The annihilation cross section was
parameterized as $\sigma_{\rm anni} = 1.2 \; \sigma_{\rm tot} /
\sqrt{s}$, where $\sqrt{s}$ is the center-of-mass energy of the
$p$-$\pb$ pair in GeV, and a power-law parameterization for the total
$p$-$\pb$ cross-section $\sigma_{\rm tot}$ from Ref.~\cite{pdg} was
used.  For consistency, we also checked the above parameterization for
anti-proton absorption loss in the TPC with the results of a full MC
simulation for the STAR TPC.  The results are consistent and within
the kinematic region of 0.4 $< p_t <$ 1.0 GeV/c, the $\pb$ absorption
loss was found to be less than 5\%.

In this paper, we report the inclusive $\pbp$ ratios and no attempts
have been made to correct for feed-down protons and anti-protons from
hyperon weak decays. Such corrections would inevitably depend on the
assumptions for hyperon and anti-hyperon yields and momentum
distributions.  However, if anti-hyperon to hyperon ratios are the
same as the $\pbp$ ratio, then weak-decay feed-downs would not affect
the $\pbp$ ratio.  Using the HIJING \cite{hijingpr} results as input,
MC simulations show that about 81\% of the $\Lambda$-decay
protons were reconstructed with $dca<3$~cm and 65\% with $dca<1$~cm.
As expected, the fractions are the same for anti-protons. For a
systematic check, some of the hyperon feed-down effects were studied
from the data by varying the $dca$ cut. The results show that the
$\pbp$ ratio decreases by about 2\% from $dca$ = 3 cm to 1 cm cut.
These effects were not included in the systematic errors estimated
below.


For this measurement, systematic errors mainly come from two sources:
(i) The systematic uncertainty in proton background in relatively low
transverse momentum region ($p_t < 0.4$ GeV/c). As mentioned above,
the background protons were estimated via the detailed GEANT
\cite{geant} simulation of the TPC. By varying the $dca$ cut, the
systematic error on the determination of number of protons is $\le$
10\% at $p_t \sim$ 0.4 GeV/c and drops to $< 2$\% at $p_t \sim$ 0.6
GeV/c.  (ii) As one will see in Figure 2(b), there is an asymmetry
between positive and negative rapidities. The asymmetry is less than
7\%.  The origin of the asymmetry is not fully understood and
therefore it is included in the overall systematic errors for the
$\pbp$ ratio. Other uncertainties like contamination in particle
identification (most at high $p_t$) and the centrality dependence of
the systematic errors are estimated much smaller, and are not included
in the overall systematic errors.

\begin{figure}[h]
\centering\mbox{
\includegraphics[width=.49\textwidth]{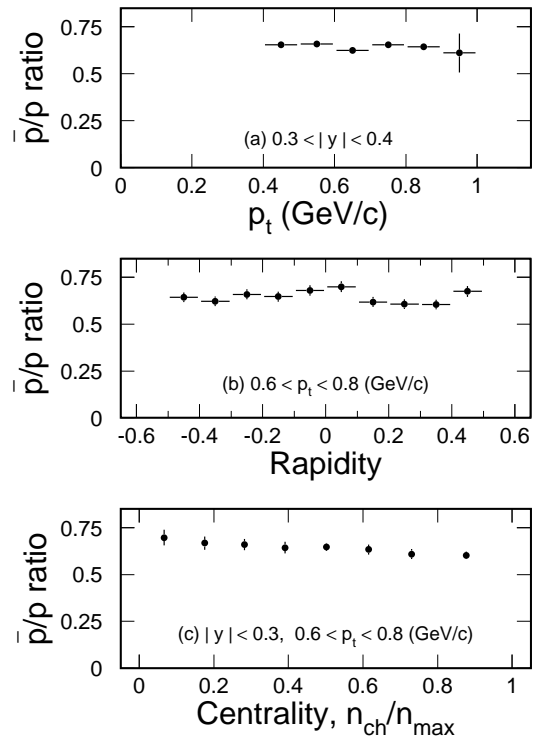}}
\vspace*{-1.75cm}
\caption{The anti-proton to proton ratios (a) as a function of 
  transverse momentum $p_t$ over the rapidity range $0.3<|y|<0.4$; (b)
  as a function of rapidity within $0.6<p_t<0.8$~GeV/$c$; and (c) as a
  function of the collision centrality within ($|y|<0.3$ and 0.6
  $<p_t<$ 0.8 GeV/$c$). (a) and (b) are for minimum bias collisions.
  Errors are statistical only. The overall systematic errors are
  estimated to be 10\%.}
\label{pbarp_yptc}
\end{figure}

Figure~\ref{pbarp_yptc}(a) shows the minimum bias $\pbp$ ratio as a
function of $p_t$ in the rapidity range of $0.3< |y| <0.4$.  The ratio
is 0.65 and is consistent with a constant value in the $p_t$ range of
$0.4 < p_t < 1.0$ GeV/$c$.  The systematic error is estimated to be
10\% for the lowest $p_t$ bin, mainly due to proton background. At
higher $p_t$ the systematic error is less than 7\%.
Figure~\ref{pbarp_yptc}(b) shows the $\pbp$ ratio integrated over
$0.6<p_t<0.8$~GeV/$c$ as a function of rapidity.  In this kinematic
region, systematic errors are estimated to be less than 10\%. Within
errors, the ratio is consistent with a constant in the measured
rapidity range.
 
Figure~\ref{pbarp_yptc}(c) shows the centrality dependence of $\pbp$
integrated over $|y|<0.3$ and $0.6<p_t<0.8$ GeV/$c$.  Although the
ratio is consistent with a constant, there is an indication of a
systematic drop from peripheral to central collisions.  At lower
bombarding energies, the value of the $\pbp$ ratio decreases by a
factor of 2 with increasing centrality for Au+Au collisions at the
AGS~\cite{e866pbarp,e814proton} and a factor of 1.6 for Pb+Pb
collisions at the SPS
\cite{na49netproton,na44proton,na44pbarp,na49pbarp,glenn}.  These low
energy results are consistent with more baryon stopping and/or
nucleon-$\overline{p}$ annihilation in central collisions relative to
peripheral collisions. A similar picture may also apply to the RHIC
results presented here.

\vspace*{-1.25cm}

At this RHIC energy, the $\pbp$ ratio is significantly smaller than
unity over the measured centrality range, indicating an overall excess
of protons over anti-protons in the mid-rapidity region. This implies
that a certain fraction of the baryon number is transported from the
incoming nucleus at beam rapidity to the mid-rapidity region even in
peripheral Au+Au collisions at $\rts=130$ GeV. Thus at this RHIC
energy, the mid-rapidity region is not yet net-baryon free.

\begin{figure}[h]
\centering\mbox{
\includegraphics[width=.48\textwidth]{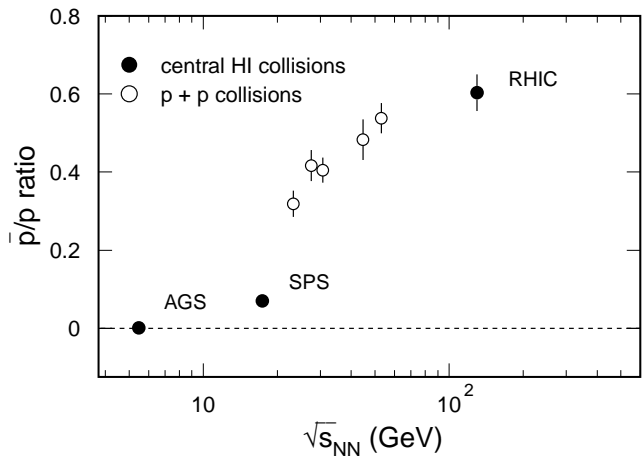}}
\caption{Mid-rapidity anti-proton to proton ratio
    ($\pbp$) measured in central heavy-ion collisions (filled symbols)
    and elementary $p+p$ collisions (open symbols).  The left end of
    the abscissa is the $p$-$\pb$ pair production threshold in $p+p$
    ($\rts=3.75$~GeV).  The RHIC data is the most central point
    from Fig. 2c. Errors, either shown or smaller than the symbol
    size, are statistical and systematic errors added in quadrature.}
\label{ecm}
\end{figure}

On the other hand, there is a dramatic increase in the mid-rapidity
$\pbp$ ratio in central heavy ion collisions in going from AGS ($\pbp
= $0.00025$\pm$10\%)~\cite{e866pbarp} to SPS ($\pbp \approx $
0.07$\pm$10\%)~\cite{na49pbarp,glenn} and to RHIC.  This is
demonstrated in Fig.~\ref{ecm} where the central heavy-ion results are
shown as a function of the center of mass energy $\sqrt{s_{\rm NN}}$.
For comparison, also shown in Fig. \ref{ecm}, are the $\pbp$ ratios in
$p+p$ collisions at mid-rapidity and averaged over $0.35 < p_t < 0.6$
GeV/$c$ \cite{ppdata}. The kinematic coverage for the $p+p$ collisions
is similar to this measurement. The value of the $\pbp$ ratio from
RHIC is close to or even larger than that in p+p collisions at
$\rts=53$~GeV.  Note that the beam rapidities at $\rts=53$ and 130~GeV
are 4.0 and 4.9, respectively.

For comparison, the HIJING(v1.35) model~\cite{hijingpr} predicts a
$\pbp$ ratio of approximately 0.8 for Au + Au central collisions at
$\rts = 130$ GeV. If the baryon junction mechanism~\cite{vance99} is
considered, the ratio is 0.75.  In contrast, the Relativistic Quantum
Molecular Dynamics RQMD(v2.4) model \cite{sorge95} predicts a $\pbp$
ratio increasing with $p_t$ with an average value of approximately
0.5. Note that the HIJING model is motivated by perturbative QCD but
does not have late stage rescattering, and the RQMD model has late
stage hadronic rescattering. As a result, the RQMD calculations
predict a $p_t$ dependence of the $\pbp$ ratio (from 0.2 to 0.55 in
$p_t \le 1$ GeV/c) while a constant ratio is observed from the HIJING
calculations.

 
In summary, we have reported the ratio of the mid-rapidity anti-proton
to proton yields in Au+Au collisions at $\rts = 130$ GeV measured by
the STAR experiment.  Within the rapidity and transverse momentum
range of $|y|<0.5$ and 0.4 $<p_t<$ 1.0 GeV/$c$, the ratio is
essentially independent of either transverse momentum or rapidity. In
this kinematic range, the average $\pbp$ ratio is found to be $0.65\pm
0.01_{\rm (stat.)} \pm 0.07_{\rm (syst.)}$ for minimum bias
collisions.  No strong centrality dependence is observed in the
centrality range measured.  The value of the $\pbp$ ratio indicates
that although $p$-$\pb$ pair production becomes important at
mid-rapidity, a significant excess of baryons over anti-baryons is
still present in heavy ion collisions at RHIC.  Comparisons of our
result to heavy ion results at lower energies indicate that the
mid-rapidity $\pbp$ ratio in heavy ion collisions increases
significantly with the collision energy.

\vspace{0.1cm}

{\bf Acknowledgments:} We wish to thank the RHIC Operations Group at
Brookhaven National Laboratory for their tremendous support and for
providing collisions for the experiment. This work was supported by
the Division of Nuclear Physics and the Division of High Energy
Physics of the Office of Science of the U.S.Department of Energy, the
United States National Science Foundation, the Bundesministerium fuer
Bildung und Forschung of Germany, the Institut National de la Physique
Nucleaire et de la Physique des Particules of France, the United
Kingdom Engineering and Physical Sciences Research Council, and the
Russian Ministry of Science and Technology

\noindent{$^{\dagger}$ Deceased.}

\end{document}